\documentclass[12pt]{article}

\usepackage{graphicx}
\usepackage{epstopdf}
\usepackage{amsmath}
\usepackage{amsfonts}
\usepackage{amssymb}
\usepackage{color}

\def\vep{\varepsilon}
\def\mpl{M_{pl}}

\def\be{\begin{equation}}
\def\ee{\end{equation}}
\def\bea{\begin{eqnarray}}
\def\eea{\end{eqnarray}}
\def\d{\partial}
\def\O{\mathcal{O}}

\newcommand{\comment}[1]{}

\begin{document}

\vspace*{0. cm}

\begin{center}

{\Large\bf  Massive Gravity: A Lorentz-Symmetric Aether}
\\[1.5 cm]
{\large Mehrdad Mirbabayi} 
\\[0.7cm]

{\normalsize { \sl  School of Natural Sciences, Institute for Advanced Study, Princeton, NJ 08540}}\\
\vspace{.2cm}

\end{center}

\vspace{.8cm}

\hrule \vspace{0.3cm}
{\small  \noindent \textbf{Abstract} \\[0.3cm]
\noindent This is a heuristic introduction to massive gravity based on an analogy with perfect fluids. I will argue that massive gravity can be thought of as Einstein gravity in the presence of a medium with unusual properties.
}
 \vspace{0.3cm}
\hrule

\begin{flushleft}
\end{flushleft}

\parskip 0.23 cm

\subsection*{Perfect fluids in Lagrangian space}

Let's start by reviewing the Lagrangian description of perfect fluids, a very well-understood class of media. The discussion closely follows that of reference \cite{Dubovsky_fluids}. Imagine labeling the elements of a fluid by a set of three scalar fields $\{\phi^1(x),\phi^2(x),\phi^3(x)\}$. This can be done, for instance, by taking $\phi^I = x^I$ at a given time $t=t_0$. The equation $\phi^I(x)=c^I$, with constant $\{c^1,c^2,c^2\}$, then implicitly gives the trajectory $x^I(t)$ of a fluid element with $x^I(t_0)=c^I$. As long as trajectories do not cross the map $\{\phi^I(x)\}$ is invertible. The fluid four-velocity $u^\mu$ can be defined by requiring
\be
u^\mu\d_\mu \phi^I=0,\qquad \text{and}\qquad u^\mu u_\mu = 1,
\ee
whose solution is 
\be
u^\mu = \frac{J^\mu}{\sqrt{J^\mu J_\mu}},
\ee
where $J^\mu$ is defined using the Levi-Civita symbol 
\be
J^\mu=\vep^{\mu\nu\alpha\beta}\d_\nu\phi^1\d_\alpha\phi^2\d_\beta\phi^3.
\ee
Once the equation of state, which relates the pressure $p$ of the fluid to its energy density $\vep$, is specified, one can complete the dictionary between the Eulerian and Lagrangian quantities as follows. Note first that $B^{1/2}\equiv \sqrt{J^\mu J_\mu}$ gives the density of fluid particles which for an isentropic motion is proportional to the entropy density $s$. Using the hydrodynamic relation 
\be
\frac{d\vep}{\vep+p}=\frac{ds}{s},
\ee
we obtain
\be
\label{vep(B)}
\frac{d\vep}{\vep+p}=\frac{1}{2}\frac{dB}{B},
\ee
which can be integrated to give $\vep(B)$ and $p(B)$.

The scalar fields $\phi^I$ can also be used to give a variational description of the dynamics of dissipationless fluids. The fluid action can be constructed by the following considerations. One can freely shift the fields $\phi^I\to \phi^I+c^I$ by constants $c^I$, so each $\phi^I$ should be acted upon by at least one derivative. At low energies the most relevant terms are those constructed from the first derivative $\d_\mu\phi^I$. By general covariance, the action depends on $\d_\mu\phi^I$ only through the matrix
\be
\label{B}
B^{IJ}=-g^{\mu\nu}\d_\mu\phi^I\d_\mu\phi^J,
\ee
and rotational symmetry further restricts it to be a function of traces of products of $B^{IJ}$. There are three independent such rotational invariants in three spatial dimensions. However, a fluid is by definition insensitive to volume-preserving deformations; therefore, out of the three invariants only the previously defined particle density squared $B=\det(B^{IJ})$ survives. Hence, the low-energy fluid action is fully specified by one single-variable function:
\be
\label{S1}
S_{\rm fluid}=\int d^4x \sqrt{-g}F(B).
\ee

To make sure that this action actually describes perfect fluids, let us derive the tensor of stress and energy by varying $S$ with respect to the metric:
\be
T_{\mu\nu}=\frac{2}{\sqrt{-g}}\frac{\delta S}{\delta g^{\mu\nu}}=-2F'B u_\mu u_\nu+(2F'B-F)g_{\mu\nu},
\ee
where I have used $B^{-1}_{IJ}\d_\mu\phi^I\d_\nu\phi^J=-u_\mu u_\nu+g_{\mu\nu}$. This is exactly the form of the stress-energy tensor of a perfect fluid if we identify the energy density and pressure as 
\be
\vep = -F,\qquad p = F-2F'B,
\ee
in agreement with eq.\eqref{vep(B)}. The fluid dynamics is entirely governed by the conservation of the stress-energy tensor, $\nabla_\mu T^\mu_\nu=0$, which in turn follows from the equations of motion for $\phi^I$ fields. 

So much for fluids in isolation, equipped with the generally covariant description \eqref{S1}, let us now turn on gravity and consider the fluctuations of a system which in the absence of gravity would have a uniform density $B_0^{1/2}$. Using the freedom of coordinate choice, we can choose spatial coordinates to match $\phi^I$ at all times: $x^I(t)= B_0^{-1/6}\phi^I$ for any constant $B_0$. This gives
\be
S_{\rm fluid}=\int d^4x \sqrt{-g}F(-B_0\det g^{IJ}).
\ee
On the other hand, if the labeling $\phi^I$ is done in a more or less uniform way, the metric fluctuations $h_{AB}=g_{AB}-\eta_{AB}$ are guaranteed to remain small in a finite space-time region well within Jeans instability scale. (The Latin indices $A,B=0,1,2,3$ are used throughout the text to emphasize that a specific coordinate choice has been made.) Expanding the action in Taylor series around $h_{AB}=0$, we obtain
\be
\label{Sh}
S_{\rm fluid}=\int d^4x [-\vep-\frac{1}{2}(\vep h_{00}+p h_{II})+a_1 h_{00}^2+a_2 h_{0I}^2+a_3 h_{00}h_{II}+a_4 h_{IJ}^2+a_5 h_{II}^2+\O(h^3)],
\ee
where $F$ and its derivatives (and hence $\vep$ and $p$) are calculated at $B_0$, the repeated indices are summed over, and the coefficients $a_i$ are determined in terms of $F$ and its derivatives. The above expression is manifestly rotationally invariant, however, unless $F$ is a constant Lorentz symmetry is broken. For instance, the linear terms do not add into a multiple of $h\equiv \eta^{AB}h_{AB}$, and the quadratic terms do not pack into Lorentz invariant combinations $h_{AB}^2\equiv \eta^{AB}\eta^{CD}h_{AC}h_{BD}$ and $h^2$. This is of course expected since the rest-frame of the fluid defines a preferred reference system and breaks symmetry under boosts.

Another important feature of the action \eqref{Sh} is the presence of terms linear in $h_{AB}$. They indicate that $g_{AB}=\eta_{AB}$ is not a solution of the Einstein equation. The fluid carries energy and in a gravitational theory it curves the space-time. However, it may be possible that at a special value of density $B_0^{1/2}$ the fluid pressure vanishes, in which case $h_{IJ}=0$ would be a valid solution at the linear level --- if we considered an elastic medium instead of a fluid, such a pressureless equilibrium point would always exist. The terms quadratic in $h_{IJ}$ in the action can then be naturally interpreted as representing the elasticity of the medium. Since the coordinate system is glued to the medium elements, expansion and compression of the medium result in variations of the metric. Away from equilibrium, $h_{IJ}$ deviate from zero, and the quadratic terms (if positive-definite) correspond to the restoring forces that resist deformations. 

\subsection*{Lorentz symmetric aether}

Consider now the Lorentz symmetric aether, a medium whose presence does not break Lorentz symmetry and carries zero energy (and pressure) when relaxed. From the above discussion, writing the general form of its action is an easy task. By Lorentz symmetry all indices must be contracted with $\eta^{AB}$. The absence of stress and energy ensures that Minkowski space is a solution of Einstein equations, and hence the action must start quadratic in $h_{AB}$:
\be
\label{aether}
S_{\rm aether} \propto\int d^4x [h_{AB}^2-a h^2+\O(h^3)].
\ee
With a little imagination this can be called Lorentz invariant massive gravity. Despite having zero stress and energy at equilibrium, the aether exists since it resists (or at least reacts to) deformations, very much like an elastic medium. 

Stretching rubber can decrease its energy density, so once the equilibrium energy density of the aether is set to zero, it is natural to expect it to become negative upon deformations. Indeed, generically it does go below zero, this is how the Yukawa screening of the gravitational field arises in massive gravity. This, however, does not lead to a catastrophic instability of the medium as long as exciting waves costs positive energy. In fact, if we were not dealing with a theory of gravity, the zero of energy would be completely arbitrary. Nevertheless, the negativity of energy will have important consequences in the study of black holes.\footnote{It was argued in \cite{discharge} that black holes disappear in massive gravity, an analogous phenomenon to the well-known process of black hole discharge in massive electrodynamics.}

On the other hand, requiring the excitation energies to be positive uniquely fixes the coefficient $a$ in eq.\eqref{aether}. Fierz and Pauli showed that unless $a=1$ the gravitational theory of aether has six degrees of freedom (six types of waves): two polarizations of a transverse traceless tensor, two polarizations of a transverse vector, and two scalars \cite{FP}. One of the scalars is a ghost, that is, its kinetic energy has an opposite sign. Consequently, the higher the frequency of excitations the more negative are their energy, leading to an instantaneous instability through coupling to positive energy modes. When $a=1$, the ghost is absent at the quadratic level and the spectrum of excitations coincides with the expectation from the representation theory of Poincar\'{e} group for a massive spin-2 particle. 

One way to see why the Fierz-Pauli choice $a=1$ is special is by comparison with the canonical formulation of the Einstein gravity \`{a} la Arnowitt, Deser, and Misner \cite{ADM}. In that formalism one first decomposes the metric into the spatial part $^{(3)}\!g_{ij}=-g_{ij}$, and the temporal components: the lapse $N=1/\sqrt{-g^{00}}$, and the shift vector $N^i= \; ^{(3)}\!g^{ij}g_{0j}$. One then observes that there is no time derivative acting on $N$ and $N^i$ in the action. Moreover, in the canonical formalism (when working with the Hamiltonian) $N$ and $N^i$ appear linearly, and hence provide 4 constraints. These constraints reduce the 6 a priori degrees of freedom in the spatial part of the metric $g_{ij}$ to 2 polarizations of the massless graviton. 

In Lorentz invariant massive gravity, $N^i$ always appears nonlinearly and the three associated constraints will be lost. As for the lapse, the Fierz-Pauli mass term is the unique quadratic expression that is linear in $\delta N=N-1$, and the associated constraint removes the above-mentioned ghost, resulting in five dynamical degrees of freedom. 

Hence, up to the quadratic level the theory is completely fixed except for an overall coefficient that determines the graviton mass. However, in a gravitational theory there is no reason for $h_{AB}$ to remain small -- both close to gravitating objects or at cosmological distances there can be order one deviations of the metric from $\eta_{AB}$. The higher order terms in eq.\eqref{aether} are, therefore, equally important, and one has to consider Fierz-Paulian massive gravity (FP) as the class of theories described by the Einstein-Hilbert action plus the nonlinear version of the Fierz-Pauli action:
\bea
\label{S}
&S=S_{EH}+S_{FP}, &\\
&S_{EH}=-{1\over 2}\mpl^2\int d^4x\sqrt{-g}R,& \\
\label{FP}
&S_{FP}=m^2\mpl^2\int d^4x\sqrt{-g}U.&
\eea 
Here $\mpl =1/ 8\pi G $, and $U$ is a Lorentz-invariant potential defined using a flat reference metric $\eta_{AB}$. For small $h_{AB}=g_{AB}-\eta_{AB}$ it reduces to the Fierz-Pauli mass term 
\bea
\label{FP1}
U^{(2)}= \frac{1}{8}(h^2-h_{AB}^2).
\eea
where $h=\eta^{AB}h_{AB}$. The Lorentz invariance of $U$ can be enforced by requiring that $U$ is a symmetric function of the eigenvalues of the matrix $H^A_B=g^{AC}\eta_{BC}$. 

Rather obviously in the hindsight, for a generic $U$ the FP action will not remain linear in $N$ beyond quadratic level, as Boulware and Deser pointed out \cite{BD}. This results in a sixth degree of freedom which they showed to be a ghost. This problem has been solved by de Rham, Gabadadze, and Tolley \cite{dRGT} who found a particular two-parameter family of potentials which propagates just five degrees of freedom at all orders. A concise representation of the family is \cite{Nieuwenhuizen}
\be \label{FP2}
U=\sum \lambda _A\lambda_B+c_2\sum \lambda _A\lambda _B\lambda _C+{c}_3\lambda _0\lambda _1\lambda _2\lambda _3,
\ee
where the sums are over all all-distinct pairs and triples of indices, and $\lambda _A$ are the four eigenvalues of the matrix
\be
\label{1-H}
\delta ^A_B - \sqrt{ H^A_B }.
\ee
It is easy to understand why this two-parameter family, which I henceforth refer to as dRGT, is special. Restricting to metrics with zero shift vector, the lapse appears only in $\lambda_0=1-N^{-1}$. Apparently, the expression \eqref{FP2} is the only symmetric combination of $\lambda_A$, such that $\sqrt{-g}U_{dRGT}$ has zero cosmological constant, and is linear in $N$. The full proof of the absence of Boulware-Deser ghost is given by Hassan and Rosen \cite{Hassan}.

As formulated above, the FP theory is not generally covariant. However, it can be made so by introducing 4 non-canonical scalar fields \cite{Siegel,AGS,Dubovsky} to write $H^A_B$ as 
\be
\label{H}
H^A_B=\eta _{BC}g^{\mu \nu}\partial _\mu\phi ^A\partial _\nu\phi ^C.
\ee
In the so-called unitary gauge, one uses these scalar fields as coordinates, $X^A=\phi^A$, and recovers the original formulation. 

Recalling our initial treatment of fluids [and in particular comparing with eq.\eqref{B}], we clearly recognize $\phi^A$ as the Lagrangian labeling of the elements of the aether. The price of having a Lorentz symmetric medium has been to introduce four instead of three scalar fields. On the other hand, in a generally covariant theory one expects to get two tensor modes from the metric which in addition to the four scalar fields give, naively, a total of six degrees of freedom. This exceeds what is needed for a massive spin-2 graviton by one. Therefore, somewhat miraculously one combination of the four scalar fields $\phi^A$ must be non-dynamical. This is of course a reformulation of the above statement about the linearity of the ADM lapse function and the associated constraint. The main goal of \cite{proof} was to study the same question using the Lagrangian description of aether.\footnote{The advantage of the approach in \cite{proof} is that it reveals useful information about the spectrum of fluctuations and their properties around different background solutions by identifying various helicity degrees of freedom. This identification had been known around the Minkowski background $g_{AB}=\eta_{AB}$ and a subclass of background solutions which are close to a scalar diffeomorphism of Minkowski $g_{AB}=\eta_{AB}+\d_A\d_B\zeta$, after the seminal work of Arkani-Hamed, Georgi, and Schwartz \cite{AGS}. The approach presented in \cite{proof} generalizes that to other backgrounds.}

It should be quite clear from the nonlinear form of the FP action that whenever the curvature scale of the background, $l$, becomes shorter than the graviton Compton wavelength $m^{-1}$, there will be large (and, to my knowledge, sign-indefinite) corrections to the kinetic term of vector and scalar fluctuations. So to be clear, dRGT is free from the Boulware-Deser ghost which is the sixth degree of freedom, but among the existing five degrees of freedom the scalar and the vector modes are likely to become superluminal, ghost-like, or strongly coupled when the metric $g_{AB}$ deviates from Minkowski (see, e.g., \cite{deRham_dS,Gruzinov_suplum,deRham_suplum,Deser}).

These pathologies are certainly threats -- to say the least -- to the consistency of the theory of massive gravity. In particular, superluminal propagation speed suggests the possibility of constructing background solutions with closed time-like curves, which lead to various paradoxes if they can in principle be realized and sustained for long enough time. See for instance \cite{Adams,Goon,Evslin,Burrage,Evslin_stable} for discussions regarding the existence and stability of these solutions in massive gravity and the related galilean theories. Nonetheless, this issue seems to remain rather contentious. An extensive review of various theoretical and phenomenological aspects of massive gravity can be found in \cite{deRham_review}.

\section*{Acknowledgments}

This note is an extract from my PhD thesis. I am grateful to Guido D'Amico, Lasha Berezhiani, Sergei Dubovsky, Massimo Porrati, Rachel Rosen, and specially to my thesis advisors Gregory Gabadadze and Andrei Gruzinov for several discussions and collaborations. This work was supported by NSF grants PHY-1314311 and PHY-0855425.


\begin{thebibliography}{99}\addcontentsline{toc}{chapter}{Bibliography}

\bibitem{Dubovsky_fluids} 
  S.~Dubovsky, L.~Hui, A.~Nicolis and D.~T.~Son,
  Phys.\ Rev.\ D {\bf 85}, 085029 (2012)
  [arXiv:1107.0731 [hep-th]].

\bibitem{discharge} 
  M.~Mirbabayi and A.~Gruzinov,
  arXiv:1303.2665 [hep-th].

\bibitem{FP}
  M.~Fierz and W.~Pauli,
  Proc.\ Roy.\ Soc.\ Lond.\  A {\bf 173}, 211 (1939).
  
\bibitem{ADM}
  R.~L.~Arnowitt, S.~Deser, C.~W.~Misner,
  
  [gr-qc/0405109].
  
\bibitem{BD}
  D.~G.~Boulware and S.~Deser,
  Phys.\ Rev.\  D {\bf 6}, 3368 (1972).

  
\bibitem{dRGT}
  C.~de Rham and G.~Gabadadze,
  Phys.\ Rev.\  D {\bf 82}, 044020 (2010)
  [arXiv:1007.0443 [hep-th]].
  
  C.~de Rham, G.~Gabadadze, A.~J.~Tolley,
  Phys.\ Rev.\ Lett.\  {\bf 106}, 231101 (2011).
  [arXiv:1011.1232 [hep-th]].

\bibitem{Nieuwenhuizen}
  T.~M.~Nieuwenhuizen,
  Phys.\ Rev.\  D {\bf 84}, 024038 (2011)
  [arXiv:1103.5912 [gr-qc]].

\bibitem{Hassan} 
  S.~F.~Hassan and R.~A.~Rosen,
  Phys.\ Rev.\ Lett.\  {\bf 108}, 041101 (2012)
  [arXiv:1106.3344 [hep-th]].

  S.~F.~Hassan and R.~A.~Rosen,
  JHEP {\bf 1204}, 123 (2012)
  [arXiv:1111.2070 [hep-th]].

\bibitem{Siegel} W.~Siegel,
  Phys.\ Rev.\  {\bf D49}, 4144-4153 (1994).
  [hep-th/9312117].

\bibitem{AGS}
  N.~Arkani-Hamed, H.~Georgi and M.~D.~Schwartz,
  Annals Phys.\  {\bf 305}, 96 (2003).


\bibitem{Dubovsky}
  S.~L.~Dubovsky,
  JHEP {\bf 0410}, 076 (2004).
  [hep-th/0409124].

\bibitem{proof} 
  M.~Mirbabayi,
  Phys.\ Rev.\ D {\bf 86}, 084006 (2012)
  [arXiv:1112.1435 [hep-th]].


\bibitem{deRham_dS}
  C.~de Rham, G.~Gabadadze, L.~Heisenberg, D.~Pirtskhalava,
  Phys.\ Rev.\  {\bf D83}, 103516 (2011).
  [arXiv:1010.1780 [hep-th]].


\bibitem{Gruzinov_suplum}
  A.~Gruzinov,
  [arXiv:1106.3972 [hep-th]].

\bibitem{deRham_suplum}
  C.~de Rham, G.~Gabadadze, A.~J.~Tolley,
  [arXiv:1107.0710 [hep-th]].

\bibitem{Deser} 
  S.~Deser and A.~Waldron,
  Phys.\  Rev.\  Lett.\  110, {\bf 111101} (2013)
  [arXiv:1212.5835 [hep-th]].

\bibitem{Adams} 
  A.~Adams, N.~Arkani-Hamed, S.~Dubovsky, A.~Nicolis and R.~Rattazzi,
  JHEP {\bf 0610}, 014 (2006)
  [hep-th/0602178].

\bibitem{Goon} 
  G.~L.~Goon, K.~Hinterbichler and M.~Trodden,
  Phys.\ Rev.\ D {\bf 83}, 085015 (2011)
  [arXiv:1008.4580 [hep-th]].


\bibitem{Evslin} 
  J.~Evslin and T.~Qiu,
  JHEP {\bf 1111}, 032 (2011)
  [arXiv:1106.0570 [hep-th]].

\bibitem{Burrage} 
  C.~Burrage, C.~de Rham, L.~Heisenberg and A.~J.~Tolley,
  JCAP {\bf 1207}, 004 (2012)
  [arXiv:1111.5549 [hep-th]].

\bibitem{Evslin_stable} 
  J.~Evslin,
  JHEP {\bf 1203}, 009 (2012)
  [arXiv:1112.1349 [hep-th]].

\bibitem{deRham_review} 
  C.~de Rham,
  Living Rev.\ Rel.\  {\bf 17}, 7 (2014)
  [arXiv:1401.4173 [hep-th]].

\end{thebibliography}
\end{document}